\documentclass{elsart}
\usepackage{times}
\usepackage{graphicx}   
\usepackage{wrapfig,psfig,epsf}  
\newcommand{\dmsq}{\Delta m^2}

\begin{document}

\begin{frontmatter}
\title{Horizontal Muons and a Search for \\AGN Neutrinos in Soudan~2 } \vskip -20pt
\author[minn]{D. DeMuth},
\author[ral]{G. J. Alner},
\author[anl]{D. Ayres},
\author[anl]{W. L. Barrett},  
\author[minn]{P. Border},
\author[ral]{D. J. A. Cockerill},
\author[ox]{J. H. Cobb},
\author[minn]{H. Courant}, 
\author[anl]{T. Fields}, 
\author[tufts]{H. Gallagher}, 
\author[anl]{M. C. Goodman},
\author[minn]{R. Gran},
\author[anl]{T. Joffe-Minor},
\author[tufts]{T. Kafka},
\author[minn]{S. Kasahara},
\author[minn]{P. J. Litchfield},
\author[tufts]{W. A. Mann},
\author[minn]{M. Marshak},
\author[tufts]{R. Milburn},
\author[minn]{W. Miller},
\author[minn]{L. Mualem},
\author[tufts]{A. Napier},
\author[tufts]{W. Oliver},
\author[ral]{G. F. Pearce},
\author[minn]{E. Peterson},
\author[minn]{D. Petyt},
\author[minn]{K. Ruddick},
\author[tufts]{M. Sanchez}
\author[tufts]{J. Schneps},
\author[tufts]{A. Sousa},
\author[minn]{B. Speakman},
\author[anl]{J. Thron},
\author[anl]{H.J. Trost},
\author[anl]{J. Uretsky},
\author[minn]{G. Villaume},
\author[ox]{N. West}
\address[anl] {Argonne National Laboratory, Argonne IL, USA}
\address[minn] {University of Minnesota, Minneapolis MN, USA}
\address[ox] {University of Oxford, Oxford, Oxon, UK}
\address[ral] {Rutherford Appleton Laboratory, Didcot, Oxon, UK}
\address[tufts] {Tufts University, Medford MA, USA}

\maketitle
\vskip -10pt
\begin{abstract} We measure the 
horizontal ($|\cos(\theta_z)|<0.14$ 
corresponding to a slant depth cut 14 kmwe)
neutrino-induced muon flux ($E_\mu~>~1.8~GeV$) in Soudan 2 to be 
$4.01\pm 0.50\pm 0.30\times 10^{-13}~$cm$^{-2}$sr$^{-1}$s$^{-1}$.
From the absence of horizontal muons with  large energy loss,
we set
a limit on the flux of muon neutrinos from Active Galactic Nuclei.
\end{abstract}
\end{frontmatter}

\section{Introduction} \label{sec:intro}
\vskip -20pt
The Earth provides an effective shield against the muon
component of cosmic ray showers for overburdens exceeding 14,000
meters-water-equivalent (mwe).  Consequently, near the 
horizontal direction ($|\cos(\theta_z)|<0.14$)
at the site of the Soudan~2 detector (vertical
depth 2090 mwe), the
muon flux
is comprised almost
entirely of $\mu^{\pm}$ initiated by charged current interactions
of atmospheric muon neutrinos in the rock surrounding the detector.
In addition, astrophysical neutrino sources
may contribute to the (horizontal) muon flux.
In particular, a high energy ($>$ 5 TeV) 
contribution may arise as the result of neutrinos which
originate within astrophysical sources such as Active Galactic Nuclei 
(AGNs)\cite{r:agn1,r:agn2,r:agn3,r:agn4,r:agn5,r:agn6,r:agn7}.

Active galaxies are those with extreme variations in brightness
corresponding to abnormal emission of large amounts of energy
at optical and/or radio wavelengths.  
Colliding and exploding galaxies also fall into this
category. The nuclei of these galaxies may produce 
both neutral and charged
particles. Most of the particles either decay 
or are absorbed before they
can escape the galaxy,
 leaving only photons and neutrinos to
 be detected on Earth. Several models  
predict substantial high-energy neutrino production
from 
AGNs\cite{r:agn1,r:agn2,r:agn3,r:agn4,r:agn5,r:agn6,r:agn7},
however
there is
a theoretical limit which conflicts with many of these
models\cite{bib:Waxman}.  AGNs may collectively
give rise to a diffuse neutrino
flux, or they may be detectable as individual point sources.
Gamma Ray Burst sources have also been proposed as
another possible diffuse source of high energy 
neutrinos\cite{bib:paolis}.
\par
Charged current interactions of multi-TeV muon
neutrinos ($\nu_\mu + \bar{\nu}_\mu$) in the rock surrounding 
the Soudan~2 detector can be expected to give rise to TeV
muons at the detector.  Muons with these large energies will very
frequently exhibit catastrophic energy loss by bremsstrahlung or
pair production as they
traverse the tracking calorimeter.  This is highly improbable for
muons ($<E_\mu>\sim20~$GeV at the detector) initiated by
atmospheric neutrino reactions in the surrounding rock.  Thus it
is possible to measure the rate of high energy AGN neutrinos.
\par
In this paper we report a measurement of the 
horizontal muon flux of
$4.01\pm 0.50\pm 0.30\times 10^{-13}~$cm$^{-2}$sr$^{-1}$s$^{-1}$
and a negative search for high energy muons from AGNs.

\section{The Soudan~2 Detector} \label{sec:detector} \vskip -20pt
\begin{wrapfigure}[14]{r}{9.2cm}\centerline{\psfig
{file=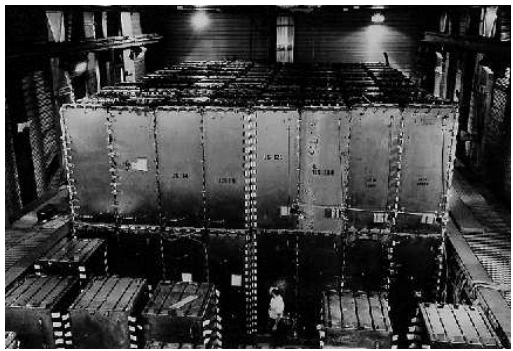,width=190pt,bbllx=90bp,bblly=250bp,bburx=520bp,bbury=545bp}}
\caption{\label{fig:detector}The Soudan~2 iron tracking calorimeter is a 
8~m $\times$ 5~m $\times$ 16~m assembly of standard-unit 4.3 t modules.  Each module contains corrugated Fe sheets interleaved with  drift tubes and stacked into a fine-grained hexagonal lattice.}\end{wrapfigure}
The detector 
is located approximately 700~m below the Earth's surface 
in an historic iron mine in Soudan, Minnesota, USA.
A photograph is shown is Figure \ref{fig:detector}.

The central detector is an iron sampling calorimeter using 
drift tubes filled with 85\% Ar - 15\% CO$_2$ gas as the
active medium. The detector is made of 224 modules with dimensions of
$2.7\times1\times1.1~$m$^3$. Each module has a mass of 
$4.3~$ t. 
The modules are stacked 2 high for a total height of 5.4~m, 
and 8 wide for a
width of 8~m. The average density is $1.6~$g/cm$^3$.

\par
The drift tubes are oriented along the North-South direction
and are operated in proportional mode.  Each end 
of the tubes
is read out by an anode wire and a cathode pad which establish the 
{\it x} (East-West) and 
{\it y} coordinates of the hits. Off line, hits 
are matched in each half tube,
using pulse 
height to match the cathode and anode pulses. The {\it z} 
(North-South) position of the hit is
determined by correlating the drift times.
\par
The ceiling, floor, and walls of the detector hall 
are covered with an active shield consisting of
a two layer aluminum proportional tube array.
The
shield array resolves  charged particle crossings (``hits")
to within four microseconds and within rectangular surfaces of
10 cm $\times$ 5 m.
For most Soudan~2 analyses,
the shield is used to veto 
events in the main detector which were initiated 
by particles interacting in the surrounding rock. Use of 
coincident in-time shield hits in the present analysis is 
described in Section \ref{sec:data}.

\par The trigger for recording events requires a multiplicity of 
seven anode pulses or eight cathode pulses in blocks of 16 contiguous
readout channels. For long muon tracks
this trigger is fully efficient. 

\begin{figure}[thb]
\centerline{
\psfig{file=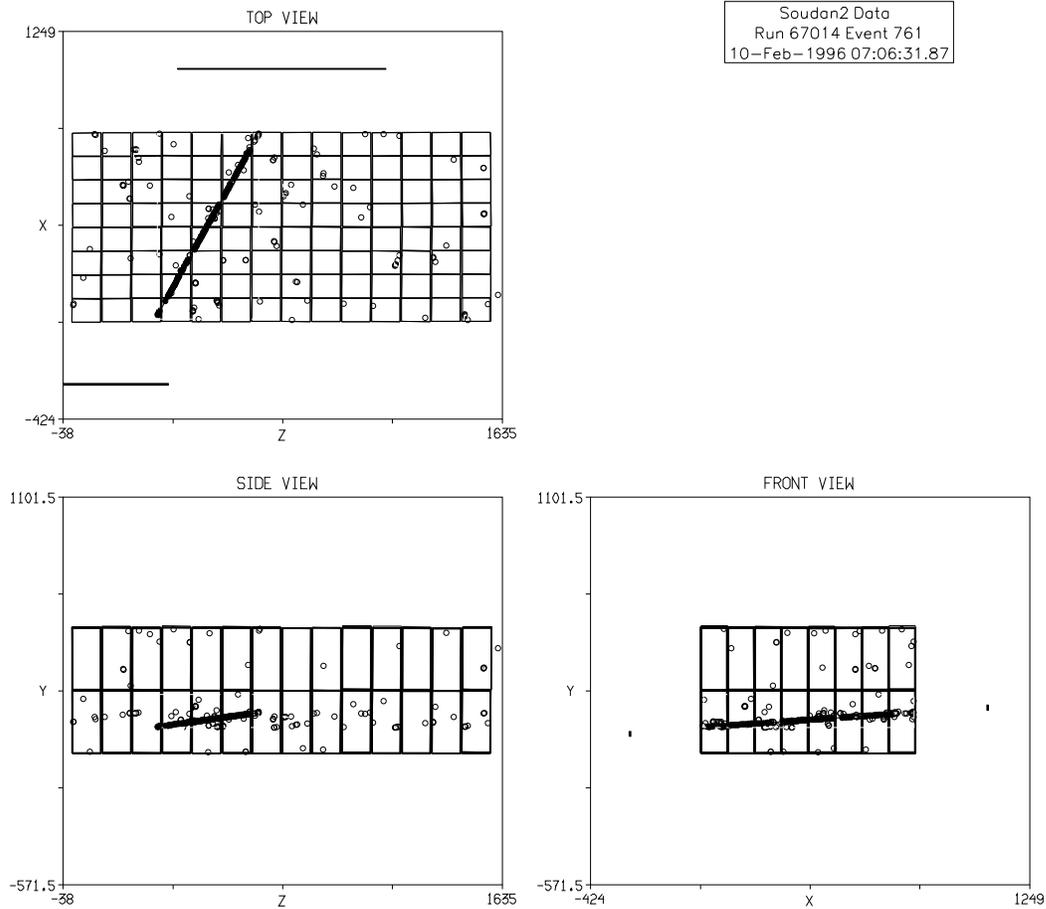,angle=90,height=5in,bbllx=40bp,bblly=170bp,bburx=560bp,bbury=740bp}}

\caption{Three views of a neutrino-induced horizontal muon 
traversing the Soudan~2 detector.  The muon traverses eight 
calorimeter modules. The proportional tube shield registers its
entrance and exit points on the cavern walls (see TOP and 
FRONT views).  Dimensions are in cm, and individual module 
outlines are shown.  All three views use the same scale so that a correct
aspect ratio is maintained.}
\label{fig:event}
\end{figure}

\par A near horizontal muon data event 
which is (presumably) neutrino-induced
is shown in Figure \ref{fig:event}.  This example is the longest
horizontal muon track
that was found, with a length of 1464 cm and
149 tube crossings used in the fit.  The zenith (azimuth)
angle of this track
is $84^\circ$ ($23^\circ$).
Two shield 
hits can be seen in the top and front views.

\par The construction and performance of the tracking
calorimeter are 
extensively described in Reference \cite{s2}.  Of particular interest
to this analysis, the single track angular resolution has been measured
using the moon shadow\cite{bib:cobb}
 to be (0.3$^\circ$ $\times$ $0.3^\circ$) and the electromagnetic
energy resolution determined by counting tube crossings
has been measured in the ISIS test beam\cite{bib:carmen} to be
\begin{equation}
\frac{\Delta E}{E} = \frac{7.0}{\sqrt{E}} + 13.5\%
\end{equation}
where E is given in GeV and the second term reflects the saturation of
the number of hits.

\par

\begin{figure}[thb]
\begin{center}
\includegraphics[width=4in]{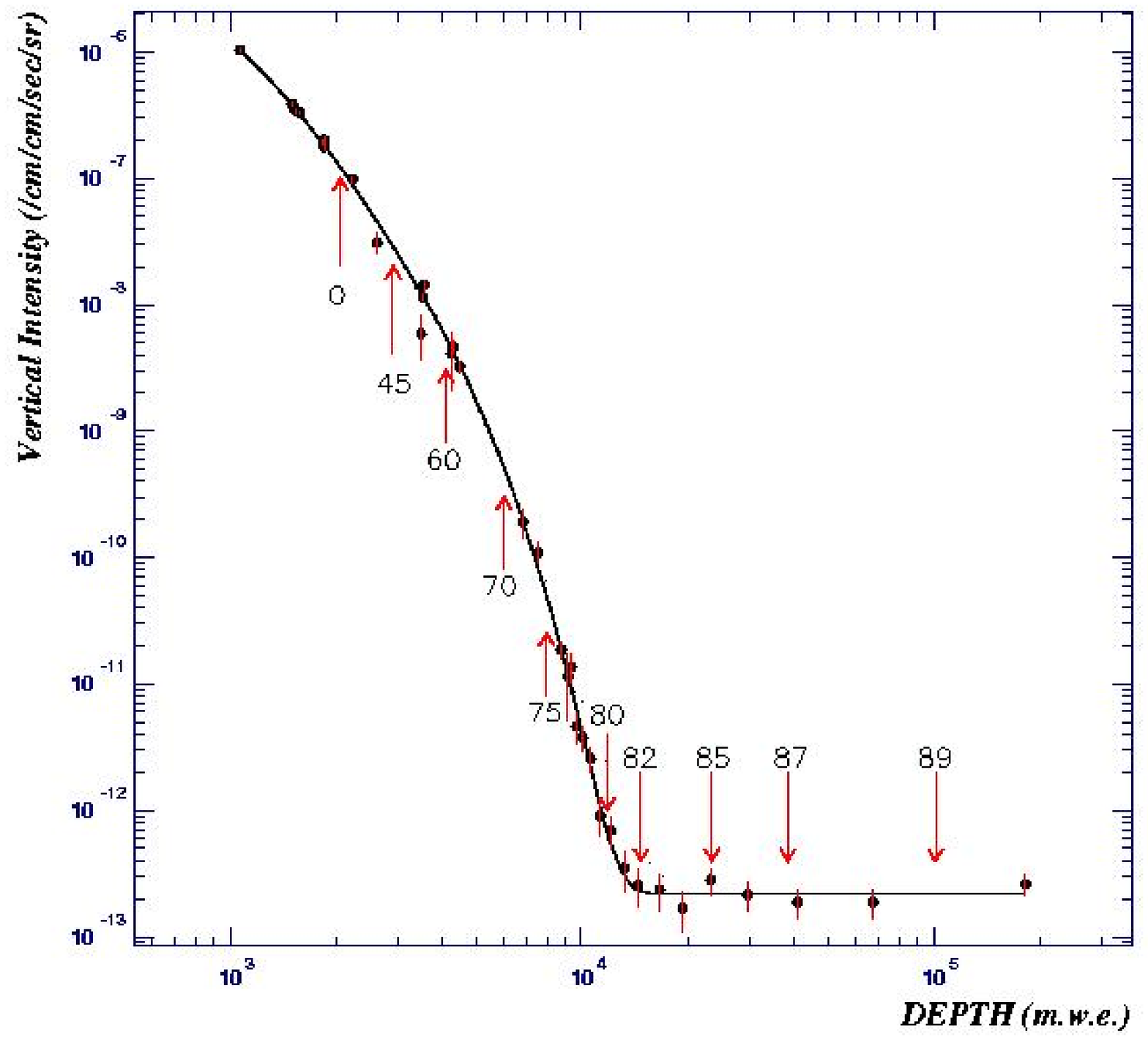}  
\caption{The vertical muon intensity 
as a function of slant depth, from the compilation
of M.F. Crouch\cite{crouch,bib:pdg}.  Arrows 
indicate the average depth at different 
zenith angles (in degrees) at the Soudan~2 detector.}
\label{fig:crouch}
\end{center}
\end{figure}
\section{Horizontal Muons at Soudan~2} \label{sec:hmu} \vskip -20pt
Figure \ref{fig:crouch} shows the vertical muon
intensity underground as a function of depth in
standard rock, as compiled by Crouch\cite{crouch}
and subsequently updated by 
the Particle Data Group\cite{bib:pdg}.
Two distinct components are apparent in the fitted
curve, which consists of a double exponential plus 
a constant term.
The 
atmospheric muon rate is observed to fall sharply with slant
depth, so that the flatter spectrum of neutrino-induced muons
only becomes visible for slant depths
greater than 14~kmwe.
For angles $\theta_z$ with respect to the vertical (zenith
angle),
the slant
depth increases approximately as $\sec(\theta_z)$, but 
depends in detail on both the surface terrain and the
rock density and composition.
\par
Based on the
vertical muon spectrum and the
calculation of the average slant depth at Soudan~2 as shown in 
Figure~\ref{fig:crouch}, we can select a sample of
neutrino induced muons by making
a cut on the zenith angle near 82$^{\circ}$.
Since the event timing resolution is about 100 ns,
it is not possible to ascertain
whether a muon is upward- or downward-going.
There is thus a two-fold ambiguity in the direction for each measured track
and $\theta_z$ is defined to be $\le 90^\circ$.

\begin{figure}[thb]
\begin{center}
\includegraphics[width=12.cm]{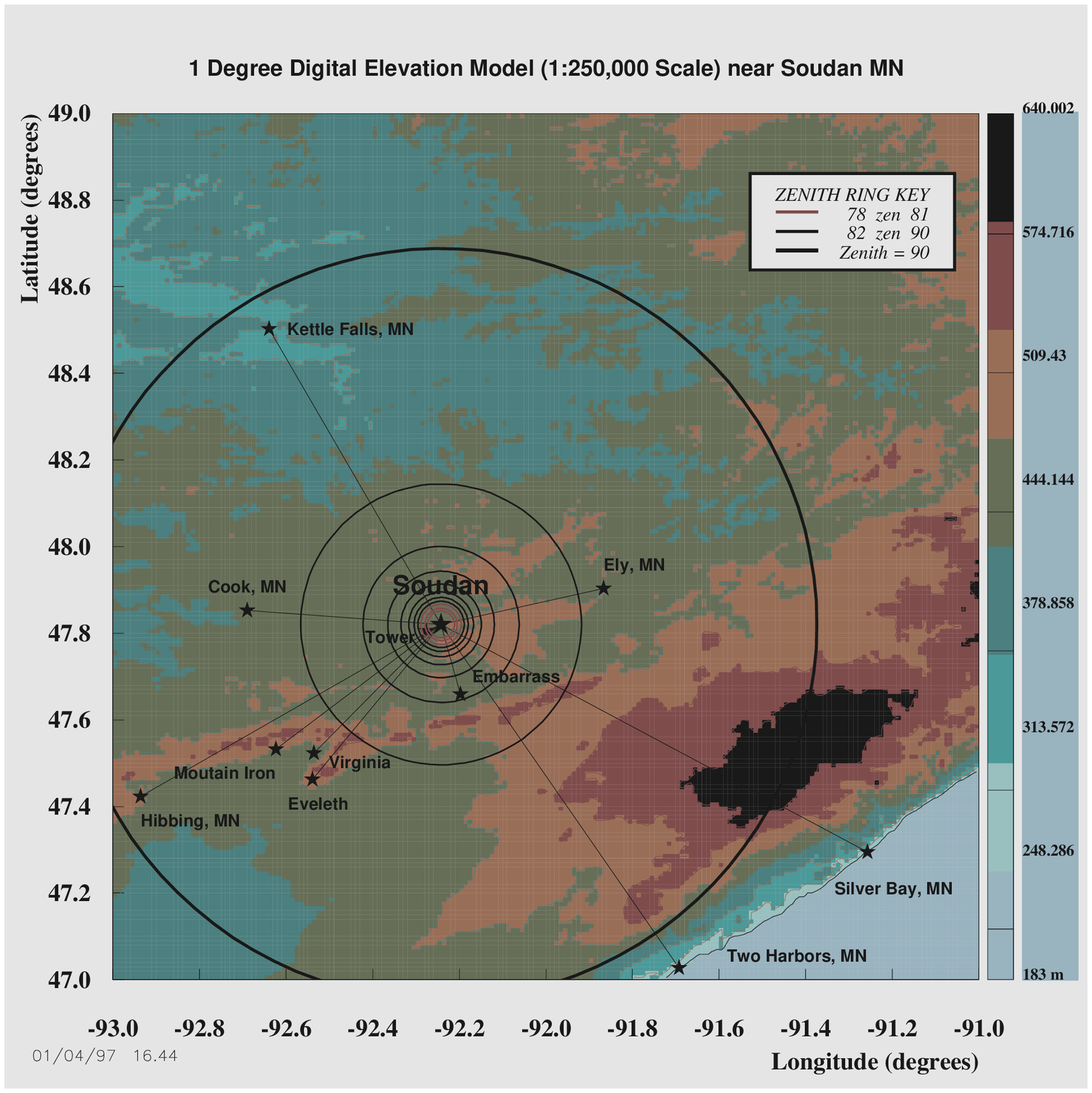}  %change back to merge.eps after edits
\vspace*{2.0mm}
\caption{USGS 1 degree digital elevation model extending 
to Lake Superior.  The zenith angles 
$78 \le \theta_z\le 90$  are projected onto the surface.  
The distance between Soudan and Two 
Harbors, MN,  is 97~km.}
\label{fig:merge}
\end{center}
\end{figure}

\par Calculation of the slant depth requires knowledge of
both the
rock density and the surface terrain.
Differences between standard rock ($\rho \sim 2.62 ~$g/cm$^3$)
and Soudan rock  (3.00~g/cm$^3~ > ~\rho > 2.74 ~$g/cm$^3$) must be taken 
into account.
The topology above and surrounding the
Soudan site is shown in Figure \ref{fig:merge}\cite{bib:usgs}.
The geology of the region is such that the iron 
deposits are almost vertical, so that
the most vertical zenith angles correspond to the highest average
density.
The majority of the rock is greenstone composed of silicon
dioxide (SiO$_2$) and aluminum oxide (Al$_2$O$_2$).\cite{bib:ely}
Based on Soudan rock effective parameters, a depth of 14 kmwe in
standard rock corresponds to 13.2 kmwe in Soudan rock.\cite{bib:demuth}

Due to the uneven surface terrain, the slant depth was
calculated for each track individually.  
Observed fluxes of cosmic ray induced muons with 
$\theta_z~<~60^\circ$ 
have been used to determine
the zenith angle dependence of the rock density\cite{bib:rud1}.
By considering the azimuthal variation in the slant depth,
we infer that systematic uncertainties in the atmospheric
muon flux 
due to incomplete knowledge of the 
rock density are on the order of 10\%.

\begin{wrapfigure}[20]{r}{9.0cm}
\centerline{\psfig{file=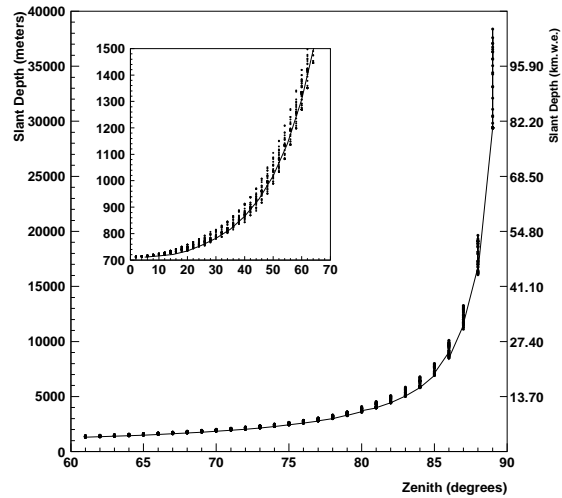,width=2.8in,bbllx=25bp,bblly=125bp,bburx=560bp,bbury=670bp}}
\caption{Estimation of slant depth versus zenith 
angle for a spherical Earth (curve) and 
upon inclusion of terrain and density information for
the Soudan site (dots). The slant depth is shown in units of meters (left)
and overburden (right).}\label{fig:s2_slantdepth}
\end{wrapfigure}

Upon inclusion of the terrain information,
a more refined estimation of the slant depth as a 
function of zenith angle is obtained, as 
indicated in Figure \ref{fig:s2_slantdepth}.
The solid curve in that Figure was made assuming a spherical Earth.
The dots which appear each one degree in zenith (two degrees in the
inset) and are averaged
over ten degrees in azimuth
reflect the effect of topology variations in the overburden.

\begin{table}[htb]
\begin{center}
\begin{tabular}{|c|l|} \hline
   & {Data Selection Criteria}\cr \hline
 1 & Track Status (extra-fiducial detector hits and in-time
shield hits)\cr
 2 & Track Length Cut ($>175$~cm)\cr 
 3 & Azimuthal Cut ($> 8^\circ$ from quadrant axis)\cr
 4 & Zenith Cut  ($\ge 78^\circ$) \cr 
 5 & Anode Width \cr \hline
\end{tabular}
\caption{Selections used to define the horizontal muon sample.}
\label{tab:cuts}
\end{center}
\end{table}

\par
\section{Reconstruction of Neutrino-Induced Muon Tracks} 
\label{sec:data} \vskip -20pt
There were
approximately $10^8$ triggered events in the
Soudan~2 detector in the data set 
considered for this analysis.  About half of these were due to cosmic
ray muons, and half due to radioactivity in different parts of 
the detector which satisfied the trigger due to multiplexing.
These events were processed
using a pattern recognition filter code
which identifies muon tracks and 
discriminates against unphysical background from 
noise\cite{bib:demuth}.  Event processing took place in three stages.
Runs were processed by the mine computer soon after the runs
were complete, and sorted into a number of output streams, one of which
was a file consisting of events which included a track
with a zenith angle greater than 60 degrees.  Those files were then
subjected to a filter program described below which identified horizontal
muon candidates with a zenith angle greater than 78 degrees.  There were
4458 events from the output of that filter which were scanned.

The filter program used five cuts
which were applied to define a ``horizontal muon" sample as listed in
Table~\ref{tab:cuts}.
Tracks were required to have end points consistent with a
particle entering and leaving 
the detector.
Acceptable track
ends were required to 1) lie outside the fiducial volume which is
50 cm from the edge of the main 
detector, or 2) lie on a module boundary inside
of the main detector and 3) point to
a coincident shield hit or 4) point to a portion of the shield where
a shield module did not exist.  
The azimuth cut 
rejected
noisy events reconstructed parallel to 
the axes of the detector.  The
zenith cut defined the sample of horizontal muons.
The track
length cut minimized short tracks such as pions that originated from
nuclear interactions within the rock nearby and also low energy
cosmic ray 
muons which had undergone large multiple Coulomb scattering.  

\par The anode-width
cut rejected events that were primarily along one anode wire.  
These were actually along the north-south direction, but
were often mis-reconstructed.  
The efficiency of these cuts for Monte Carlo muons generated
isotropically
with zenith angles exceeding 82 degrees 
is 0.56 $\pm$ 0.01 as given in Table~\ref{tab:efficiency}.
Note that both the trigger and reconstruction efficiency are
lower near zenith angles of 90 degrees.

\par
The pattern recognition software found single horizontal muon
tracks, but it also kept a variety of background events.
Most background events contained short horizontal tracks 
accompanied by  other activity in the detector.
Physicist scanning of all tracks greater than 78$^\circ$ 
was done
to verify the track fits and
to eliminate obvious backgrounds.  Common backgrounds were
due to certain patterns of electronic noise,
vertical muons undergoing large radiative stochastic losses, multiple
muons, and neutrino interactions within the detector.  
Tracks with visible multiple
scattering (greater than 2 degrees) or with in-time shield hits
that were not due to a horizontal muon were also rejected.
All
events were subjected to two independent scans
and residual discrepancies were
rescanned.
The uncorrelated scan inefficiency as determined from the double scan
was less than 1\%.  The Monte Carlo which was used for the
efficiency calculation only generated high energy (100 GeV) muons.
This fails to take into account neutrino induced muons with lower energies
in the detector which might
fail the multiple scattering cut applied during scanning.  An independent
calculation that took into account the neutrino induced muon energy
distribution at the detector gave the result that 5\% $\pm$ 5\% of the
events should undergo multiple scattering by
 more than 2 degrees in the detector, the 
criterion used during scanning.  We therefore used a scan efficiency of
95\% for the atmospheric neutrino flux calculation.  Note that this 
inefficiency does not apply to the AGN $\nu$ search in the next section.

\begin{table}[htb]
\begin{center}
\begin{tabular}{|l|r|} \hline
Cut & \\ \hline
Generate & 2000 \\ \hline
Matched Track & 1920 \\ \hline
Track Status & 1818 \\ \hline
Track Length & 1685 \\ \hline
Azimuth & 1411 \\ \hline
Zenith & 1378 \\ \hline
Anode-width & 1111 \\ \hline \hline
$\varepsilon$ & 55.6\% \\ \hline
\end{tabular}
\caption{Survival rates of 2000 Monte Carlo muons 
incident nearly horizontal on the Soudan~2 detector.}
\label{tab:efficiency}
\end{center}
\end{table}

\par The track-length cut of 1.75~m (2.2 hadronic interaction
lengths) minimized the number of
short tracks
which arose from processes which were not neutrino-related.
The latter backgrounds included pion tracks originating from deep
inelastic muon scattering within nearby rock, or else high energy cosmic
ray muons from vertical directions which experienced
large-angle multiple scattering.
This length corresponds to 
 a muon energy threshold
of $400~$MeV, but the restriction on multiple scattering raised this
to an effective threshold of $1.8~$GeV.
\par
\begin{figure}[thb]
\begin{center}
\centerline{\psfig{file=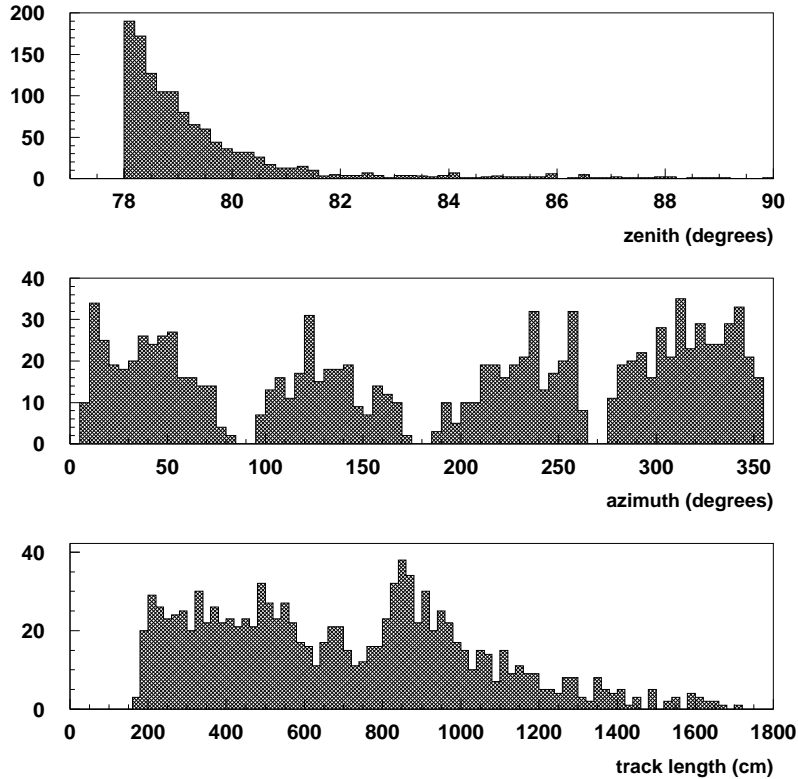,width=4.5in,bbllx=40bp,bblly=160bp,bburx=525bp,bbury=640bp}}
\caption{The reconstructed zenith, azimuth and 
track-length distributions for through-going muons 
which were selected by the filter software 
to have a zenith in excess of $78^\circ$ and
passed the scan criteria.  The shape of the track-length
distribution reflects the shape of the detector for an
isotropic muon flux.}
\label{fig:length3}
\end{center}
\end{figure}
\begin{figure}[thb]
\begin{center}
\centerline{\psfig{file=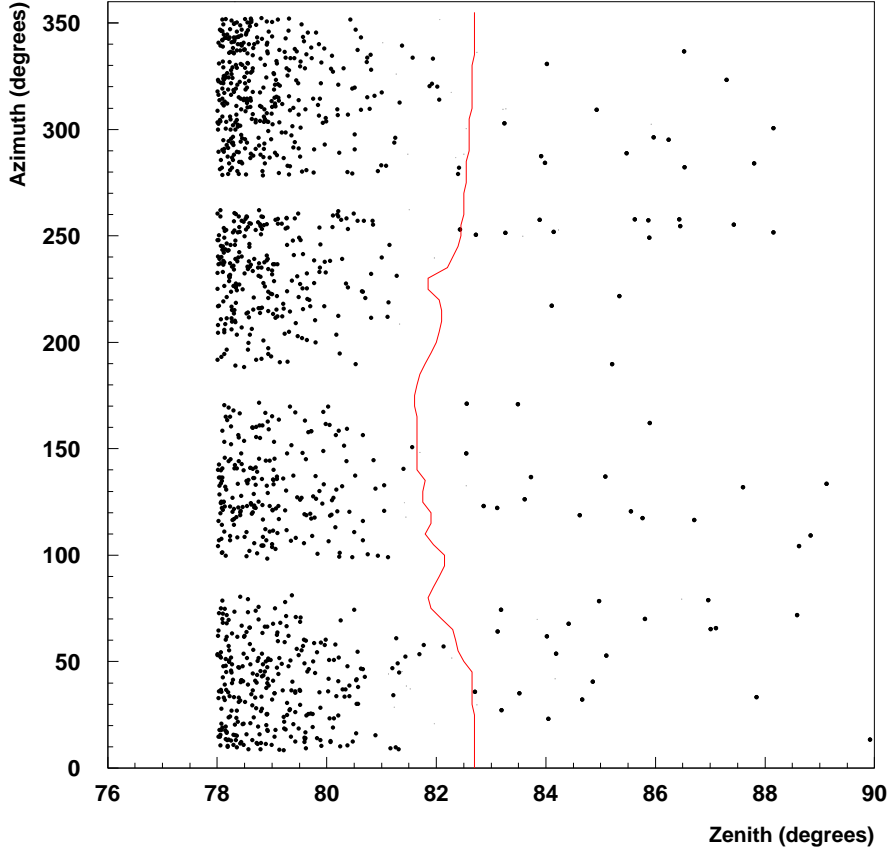,width=380pt,bbllx=40bp,bblly=160bp,bburx=560bp,bbury=680bp}}
\caption{Angular distribution of horizontal muon candidates. 
The contour line near $\theta_z~=~82^\circ$ represents 
the $14~$kmwe contour. }
\label{fig:data_theta_phi}

\end{center}
\end{figure}
\begin{figure}[thb]
\begin{center}
\centerline{\psfig{file=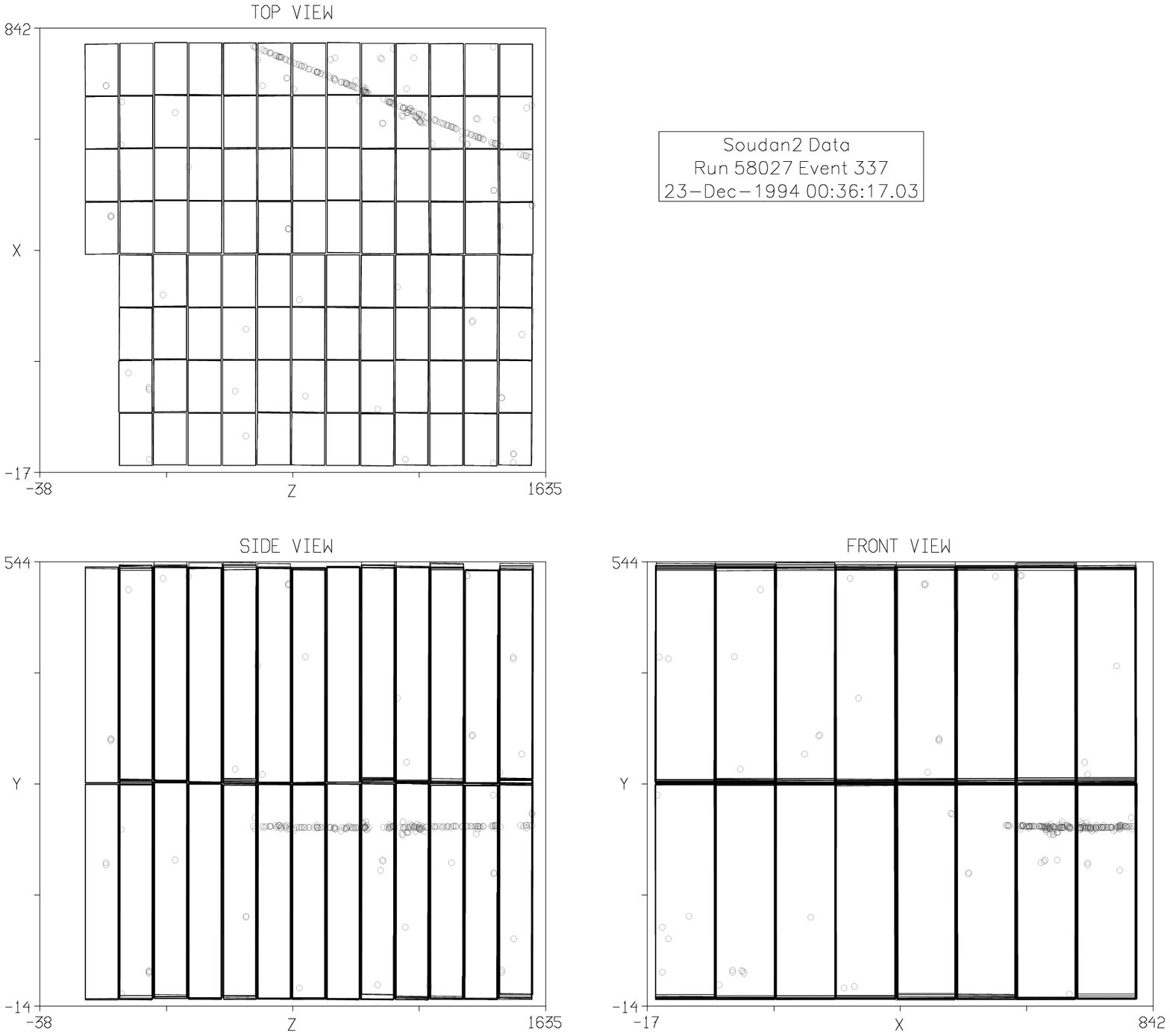,angle=90,height=5in,bbllx=40bp,bblly=170bp,bburx=560bp,bbury=740bp}}
\caption{Our largest zenith angle event. 
$\theta_z=89.9^\circ$ and the length is 731~cm.
The scales are labeled in cm.}
\label{fig:largest}
\end{center}
\end{figure}

The number of accepted events after scanning,
with $\theta_z>78^\circ$, was 1237 events.
The slant depth cut of $14~$kmwe
 reduced this sample to 65 events.
The track length, azimuth and zenith angle distributions of 
all 1237 events are shown in Figure~\ref{fig:length3}.
Figure \ref{fig:data_theta_phi} shows 
the ($\theta_z$, $\phi$) distribution of these events. 
The depletions at regular intervals on azimuthal projections
of these plots are` due to the azimuth cuts.
The wavy-line contour on Figure \ref{fig:data_theta_phi}
shows the 14~kmwe 
slant depth cut, which is designed to separate the sample
into atmospheric muon and neutrino-induced muon candidates.
The event with the largest zenith angle is shown in 
Figure~\ref{fig:largest}.


To estimate the amount of background from atmospheric muons we 
integrate under the intensity-vs.-slant-depth curve
of Figure \ref{fig:crouch}, but without the constant 
(neutrino-induced muon) term, 
for slant depths greater than $14~$kmwe.
This procedure yields a background 
estimate of $0.1^{+0.2}_{-0.1}$ events
 in our horizontal muon sample, where the error reflects the  uncertainty in
the rock density.

Muons which multiple scatter in the Earth above
the Soudan~2 detector are a potential source of additional background.
Since the overburden above the Soudan~2 detector
is so great, it is not appropriate to apply 
the usual equation for small 
angle multiple scattering.  Instead, a Monte Carlo 
calculation was performed
which propagated muons through the Earth including multiple scattering
and calculated the 
energy loss
in steps.  With a sample starting with 60 million muons that had
$\theta_z~>~60~$degrees at the surface of the Earth, the same number that 
were recorded during our exposure, no additional
muons were generated with $\theta_z~>~82~$degrees at the detector.
The finite statistics of the Monte Carlo introduced an
additional uncertainty
which corresponds to $<0.5$ event background.
We conclude that the expected background is less than one event.
\par 
The neutrino-induced muon flux, $\Phi_{\mu}$, can be expressed as
\vskip -15pt 
\begin{equation}
\Phi_{\mu}=\frac{N_{\mu}}{t\,A\,\Omega\,\varepsilon}
\end{equation}
\vskip -15pt
where $N_\mu$ is the number of observed neutrino induced muons,
t is the time exposure of the detector in seconds, A is the
effective area of the detector in cm$^2$, 
$\Omega$ is the solid angle subtended by the detector in steradians,
and $\varepsilon$ is the detection 
efficiency for the muons in the detector.  Both A and $\Omega$
are calculated assuming a uniform acceptance to the 
right of the contour in Figure \ref{fig:data_theta_phi}.

The analyzed data spanned the
period from April 14, 1992 to April 24, 2002.
The exposure time was calculated from the start and end times 
of every processed data run. 
The calculated exposure for this analysis, including corrections for the 
detector duty cycle and electronics dead time during data taking, is
{$2.00 \times10^8~$s}.  
To estimate the effective area for this analysis,
Monte Carlo muons were uniformly generated 
and the effective area for each track calculated. 
%For zenith angles greater than 
%$78^\circ$ the average area was 
%$90.6~$m$^2$. 
For $\theta_z>82^\circ$ the average effective area 
was $86.7\pm0.3~$m$^2$.
The solid angle was calculated from the acceptance region
in Figure~\ref{fig:data_theta_phi}
to be $1.77~$sr.
The trigger and reconstruction efficiency $\varepsilon$ = 0.53
(0.556 from Table \ref{tab:efficiency} $\times$  0.95 from
scanning.)
\par
Systematic errors arise with the uncertainty in the background 
of 0.7 events (1\%), uncertainties in the scan efficiency (5.6\%) and
uncertainty in the energy distribution as it affects the efficiency
of the $2^\circ$ multiple scattering cut (5\%).  Adding these 
errors in
quadrature, we assign an overall systematic uncertainty
to the neutrino flux calculation
of 7.6\%.

The resulting neutrino-induced muon flux for a muon energy threshold
of 1.8 GeV is
\vskip -15pt
\begin{equation}
\Phi_{\nu_{\mu}} = {4.01 \pm 0.50~(stat) \pm 0.30 ~(sys) 
\times 10^{-13}}\; {cm^{-2}sr^{-1}s^{-1}}
\end{equation}
\vskip -15pt

\par Upward-going atmospheric neutrino-induced muons
have been previously 
measured by Baksan, IMB, MACRO, Kamiokande, and 
Super-Kamiokande\cite{bib:bak,bib:imb,bib:macro,bib:kam,bib:sk}.  
However, these experiments have not reported
neutrino-induced muon fluxes above
 the horizon from those regions of solid angle where the overburden 
might be sufficient to separate atmospheric muons, perhaps due to
uncertainties in the surface topology and rock density.  
The SNO experiment
is deep enough that it should be able to see the neutrino-induced
muon flux above the horizon\cite{bib:SNO}.
The flux near the horizon has also
been measured by  LVD\cite{bib:lvd} and Frejus\cite{bib:daum,bib:rhode}.   
The present measurement has aspects in common with that of Frejus, 
albeit with increased statistics.  
LVD, which has a much less uniform overburden, did not
separate their neutrino-induced signal from background.  
The measured fluxes closest to $\theta_Z = 90^\circ$ are tabulated
in Table \ref{tab:fluxes} along with our estimate of the muon
energy threshold used for each analysis.
\begin{table}[!ht]
\begin{center}
\begin{tabular}{|c|c|c|c|c|} \hline
Experiment & Flux &  $E_{min}$  & $\cos \theta_z$ & Depth \\ \hline
           &  ($10^{-13} $cm$^{-2}$sr$^{-1}$s$^{-1}$) & (GeV) & &  MWE \\ \hline
Baksan\cite{bib:bak} & $4.04 \pm 2.01$ & 1    & -0.1 to 0.0 & 850 \\ \hline
Frejus\cite{bib:daum} & $3.67 \pm 0.66$ &  0.3 & -0.18 to 0.18 & 4710
\\ \hline
LVD\cite{bib:lvd} & $8.3 \pm 2.6$ &  1.0  & -0.1 to 0.1 & 3000 \\ \hline
IMB\cite{bib:imb} &  $5.66 \pm 0.95$ & 1.8  & -0.14 to 0.0 & 1570
\\ \hline
Kamiokande\cite{bib:kam} & $2.84 \pm 0.53$ & 1.7   & -0.1 to 0.0 & 2700 \\ \hline
Super-K\cite{bib:sk} & $3.45 \pm 0.33$ &  1.6  & 
-0.1 to 0.0 & 2700 \\ \hline
MACRO\cite{bib:macro} & $7.4 \pm 2.8$ &  0.4  & -0.1 to 0.0 & 3150 \\ \hline
Soudan~2 & $4.01 \pm 0.58$  &  1.8  & -0.14 to 0.14 & 2090 \\ \hline
\end{tabular} \vskip 5pt
\caption{Comparison of near-horizontal
neutrino-induced muon fluxes measured
by several experiments.  Zenith angle ranges and
energy thresholds are approximate.  Several experiments report fluxes
for restricted portions of azimuth angle to reduce backgrounds.  The
reported depths are minimum overburdens, except for Frejus and Baksan
which report effective depths.}
\label{tab:fluxes}
\end{center}
\end{table}

\par Super-Kamiokande\cite{r:osk}, and Kamiokande\cite{r:ok}, MACRO\cite{r:om},
 and Soudan~2\cite{r:os2} as well
have analyzed 
atmospheric neutrino events and concluded 
that neutrino oscillations  modify the
zenith angle distribution of the $\nu_\mu$ flavor component of
the atmospheric neutrino flux.  For a neutrino
energy of $40$~GeV and assuming $\dmsq~\sim~3.5 \times 10^{-3}~$eV$^2$
and maximal ($\theta_{23}$) mixing, 
the probability of 
oscillation for a $\nu_\mu$ from $82^\circ$ ($<L>~\sim~130~$km)
is $2.1 \times 10^{-4}$ while
for a $\nu_\mu$ from $98^\circ$ 
($<L>~\sim$ 2000 km)
it is $0.04$. 
Our modest statistics
and $90^\circ-\theta_z$ ambiguity make an oscillation analysis of
these events impractical.

\section{AGN Neutrino Search} 

Several models for neutrino production in Active Galactic Nuclei
predict large and potentially measurable fluxes of very high energy
neutrinos\cite{r:agn1,r:agn2}. These 
neutrinos 
would then produce high energy muons 
with energies from a few TeV 
up to 100s of TeV. 
In this energy range 
the dominant energy loss process
for the muons in iron is electron
pair production, followed by 
bremsstrahlung\cite{eloss}.
Both processes produce large electromagnetic showers  in
the Soudan~2 detector which are easily detected and measured.

\begin{wrapfigure}[17]{r}{7.2cm} 
\centerline{\psfig{file=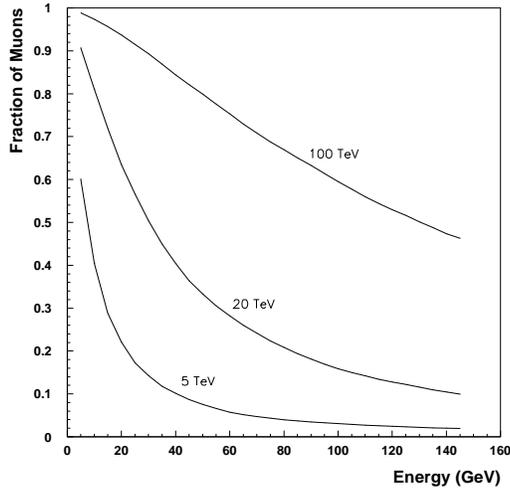,width=2.8in,bbllx=30bp,bblly=160bp,bburx=520bp,bbury=640bp}}
\caption{Calculation of non-ionization energy loss for muons in Soudan~2}
\label{fig:trost}
\end{wrapfigure}

To understand the response of the tracking calorimeter
to these TeV muons, a simulation was performed. 
High energy muons were propagated through the Soudan~2 detector and the amount 
of energy they  deposited  was calculated
in a GEANT-based Monte Carlo\cite{bib:trost}.
Three different muon energies were 
studied: 5, 20 and 100~TeV.
Figure~\ref{fig:trost} shows the fraction of muons which lost
at least
a specific amount of energy versus that energy loss for 
three different  muon energies.  The results obtained from
the simulation agree with an analytic calculation carried
out independently\cite{Uretsky}.
While 60\% of the 5~TeV muons lose 5~GeV or more, 91\% of 
the 20~TeV  muons and 99\% of the
100~TeV muons lose at least 5~GeV 
in the detector.

\begin{figure}[thb]
\centerline{\psfig{file=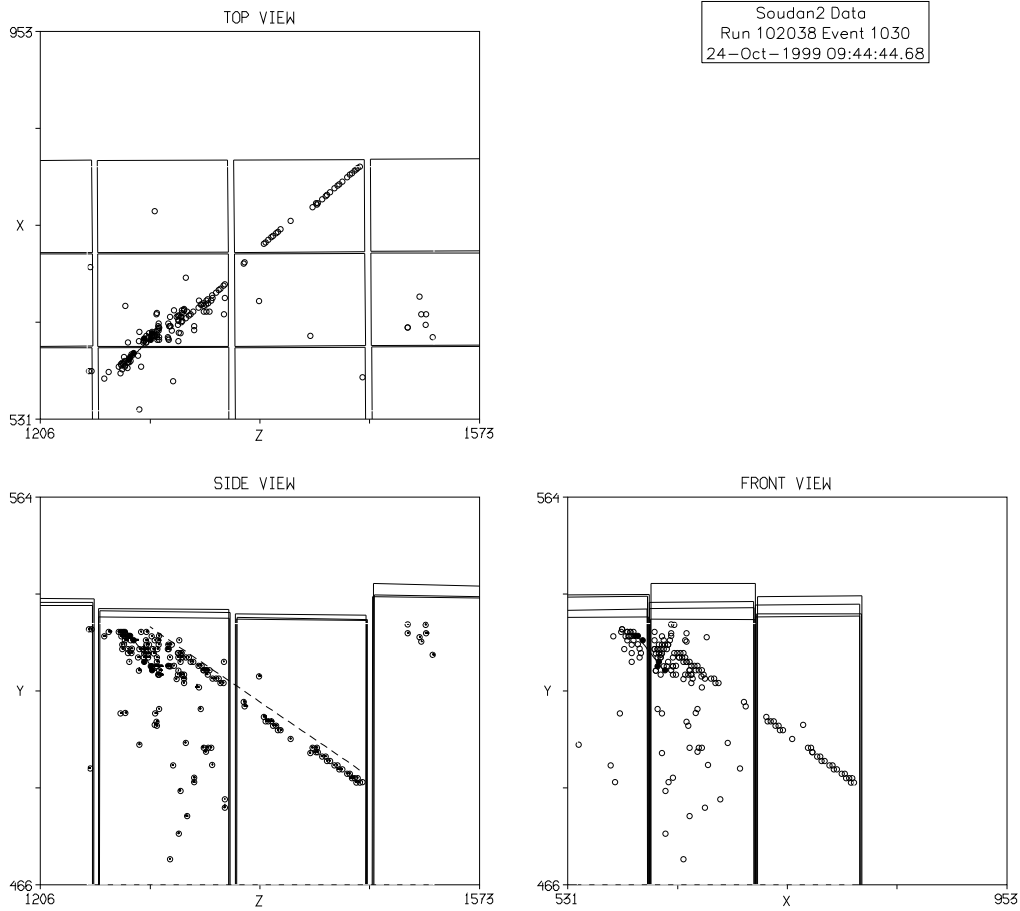,height=5in,angle=90,bbllx=40bp,bblly=170bp,bburx=560bp,bbury=740bp}}
\caption{Views of the muon event of largest
zenith angle (80.6$^\circ$) having a
substantial radiative energy  loss.
The muon's zenith angle is too small to qualify as a 
neutrino-induced candidate.  The vertical scale is magnified and
all scales are labeled in cm.}
\label{fig:run102038}
\end{figure}

The muons identified for the horizontal muon flux measurement were 
subjected to a predetermined cut of 5~GeV
on the amount of energy loss they experienced 
in the detector. None of the 65 events had visible radiated energy loss
greater than this cut.  
The largest zenith
angle event with substantial visible 
energy loss is shown in Figure \ref{fig:run102038}.
This event has a horizontal muon with a zenith angle of $80.6^\circ$
and an azimuthal angle of $133^\circ$ which entered the top of the
detector and exits the north wall.  Parallel to the muon at the
top is an entering electromagnetic shower, presumably from a
bremsstrahlung or pair production process along the muon path
in the rock
above the detector.  From Figure~\ref{fig:data_theta_phi},
the zenith angle ($80.6^\circ$)
tags the event as 
a high energy cosmic ray muon and not a horizontal muon
candidate.  Visible energy loss from a
cosmic ray muon is not unexpected, because for
atmospheric muons at the detector near zenith angle $80^\circ$, 
$<E_\mu>~\sim~380~$GeV while for atmospheric neutrino induced
muons, 
$<E_\mu>~\sim~20~$GeV.
The energy of the shower is reconstructed to be 2.2 GeV. 
There are no such events among the 65 horizontal muon candidates.
\par
Models for  neutrinos from AGNs predict a $\nu$ flux
as a function of energy.  
For each assumed 
functional form of the predicted neutrino energy spectrum, an 
observation or limit on high energy muons can be used to normalize
or limit the neutrino flux.  
It is less meaningful to plot a limit
on a neutrino energy distribution for an arbitrary functional form.
Therefore we choose to show our limit on the neutrino-induced muon
energy distribution.  For each muon energy, our non-observation of
events with energy loss corresponds to an upper limit on the muon
flux.

\par To provide a comparison of our search with other
searches reported in the literature, we show in
Table \ref{tab:agn} the
exposures ($A \times t \times \Omega \times \varepsilon$).
Our exposure is larger than that
of the Frejus experiment, but lower than MACRO or AMANDA.
Since the latter two experiments focus 
on upward-going muons, and the Frejus
and Soudan~2 limits use horizontal muons, the limits are 
complementary for the exotic case involving the highest energy 
neutrinos for which the Earth's neutrino opacity is large
($E_\nu~>~50$ TeV)\cite{bib:paolis}.
\begin{table}[!ht]
\begin{center}
\begin{tabular}{|c|c|} \hline
Experiment & Exposure ($10^{14}~$cm$^2$~s~sr)\\ \hline
Soudan~2 &  1.71  \\ \hline
Frejus\cite{bib:rhode} &  1.60  \\ \hline
MACRO\cite{bib:macroa} & 3.40  \\ \hline
AMANDA\cite{bib:amanda} &  13.9 \\ \hline
\end{tabular}
\end{center}
\caption{Comparison of exposures (Product of running time, effective area, solid angle and efficiency) for AGN neutrino search experiments}
\label{tab:agn}
\end{table}

\begin{figure}[thb]
\centerline{\psfig{file=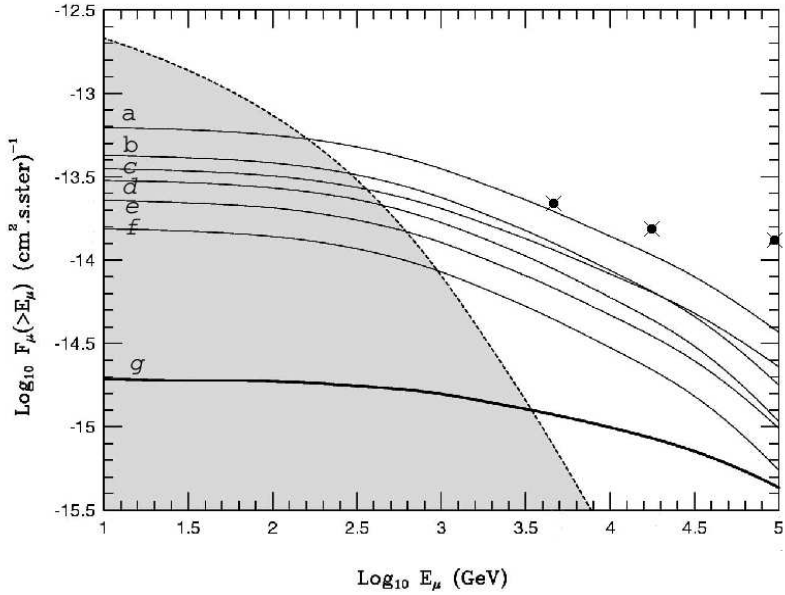,width=360pt,bbllx=16bp,bblly=16bp,bburx=680bp,bbury=520bp}}
\caption{The data points are our totally correlated limits from 
Table \ref{tab:limits}.
The Curves a-f are from horizontal muon flux predictions from models by Szabo and Protheroe \cite{r:agn2} and curve g from Stecker \cite{r:agn5}.  The shaded area represents the atmospheric neutrino flux.}
\label{fig:agn}
\end{figure}

Based upon the observation of zero events, together
with our 
efficiencies, we calculate 90\% CL upper
limits for the integral muon flux above three energies,
which
are shown in Table \ref{tab:limits} and  Figure~\ref{fig:agn}.
Our limit is close to the highest predicted flux; however that
flux is not ruled out.  The limits are lower for higher energy
cut-offs because of the higher probability of detecting a large
energy loss.

\begin{table}[htb]
\centerline{Soudan~2 Flux Limits for AGN Neutrinos}
\vskip 5pt
\begin{center}
\begin{tabular}{|c|c|c|} \hline
Energy~(TeV) & High Energy $\mu$ Efficiency & 
90\% CL Limit (cm$^{-2}$sr$^{-1}$sec$^{-1}$)\cr \hline
  5 & 60\% & $2.2 \times 10^{-14}$ \cr
 20 & 91\% & $1.5 \times 10^{-14}$ \cr
100 & 99\% & $1.4 \times 10^{-14}$ \cr \hline
\end{tabular} \vskip 5pt
\caption{Efficiency for high energy muons to experience at least 
5~GeV energy loss in the Soudan~2 detector and the resulting AGN neutrino flux limits.}
\label{tab:limits}
\end{center}
\end{table}

\par Other experiments have also used their horizontal muon events
to set limits on the flux of the diffuse AGN neutrinos. The Frejus
experiment 
used two methods to set a limit on the flux of high energy 
neutrinos\cite{bib:rhode}.  One method extrapolated a fit to the 
atmospheric neutrino energy spectrum obtained 
previously\cite{bib:daum}.  Since that fit assumed that one spectral
index fit all of their data, 
using it to deduce a limit for
muons arising from a different energy distribution may
bias that limit toward small values.
The limit was quoted as $d\phi/dE_\nu(2.6~$TeV$)
< 7.0~10^{-13} $GeV$^{-1}$cm$^{-2}$s$^{-1}$sr$^{-1}$.  With their second 
method, Frejus used the absence of muons with large energy loss to
constrain the normalization of 
a number of specific models of the high energy neutrino
flux\cite{bib:rhode}.
\par  The MACRO experiment\cite{bib:macroa} conducted an AGN 
search and found 2 candidate events with background of 1.1 expected.
They set an upper
limit on the muon flux from diffuse neutrinos
of $(1.7 \pm 0.2) \times 10^{-14}$cm$^{-2}$s$^{-1}$sr$^{-1}$.  Their
efficiency was highest for 
upward-going muons and lowest for horizontal muons.
The AMANDA neutrino
telescope, located in the ice near the South Pole,
has set a limit based on an assumed diffuse $E_\nu^{-2}$
spectrum of $dN/dE_\nu \le 
10^{-6}E_\nu^{-2}$cm$^{-2}$s$^{-1}$sr$^{-1}$GeV$^{-1}$\cite{bib:amanda}

\par
\begin{figure}[thb]
\centerline{\psfig{file=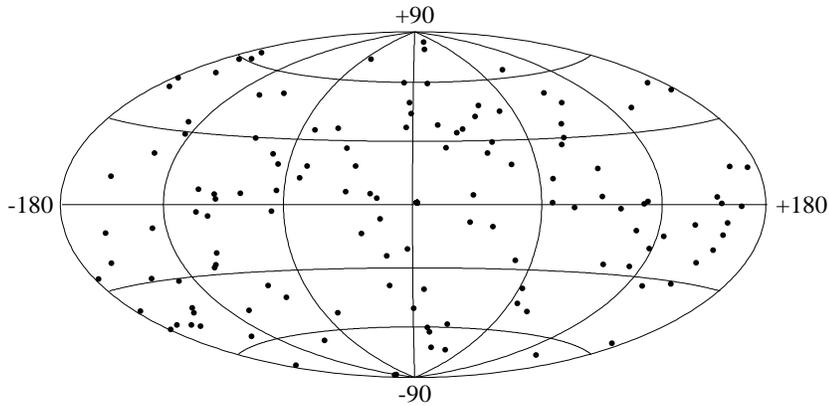,width=320pt,bbllx=50bp,bblly=300bp,bburx=560bp,bbury=560bp}}

\caption{Aitoff Projection of the 65 horizontal muon candidates.  Each event has two projections due to the ambiguity in direction. }
\label{fig:galaxy_point}
\end{figure}

As another test for an extraterrestrial component
in the horizontal muon events, we have examined them for evidence
of point sources.
Figure \ref{fig:galaxy_point} shows the directions of the 65 
horizontal muons in galactic coordinates using an Aitoff projection of the 
galaxy.
For each muon, the two possible directions are plotted.
We observe that no clusters involving 
two or more muons within the detector's angular resolution
$(0.3^\circ \times 0.3^\circ)$\cite{bib:cobb}
appear within the plot.

\section{Summary}
In this study we have
isolated a sample of 65 neutrino-induced muon candidates, with 
an estimated background of less than one 
event.  For a
slant depth cut 14 kmwe and a muon energy threshold of 1.8 GeV,
this corresponds to a muon flux of $4.01 \pm 0.50^{stat}
\pm 0.30^{sys} \times 10^{-13} $cm$^{-2}$sr$^{-1}$s$^{-1}$ in the
horizontal direction $-0.14 < \cos \theta_z <0.14$.  None of
the 65 events
have a large energy loss in the detector.  We 
set an integral
limit on neutrino-induced muons from AGNs and/or other
sources from between 1.4 and 2.2 
$\times 10^{-14} $cm$^{-2}$sr$^{-1}$s$^{-1}$ depending on muon energy.

\section{Acknowledgments} \vskip -20pt
We thank Rick Egeland, Leah M. Goodman and Adrian Ryerson
for their help in this analysis.
This work was supported by the U.S. Department of Energy, the U.K. Particle
Physics and Astronomy Research Council, and the State and University of
Minnesota.  We also wish to thank the Minnesota Department of Natural
Resources for allowing us to use the facilities of the Soudan
Underground Mine State Park.

%*************************************************************************

\end{document}